\documentclass[pre,amsmath]{revtex4-2}
\usepackage{graphicx}
\usepackage{color}
\usepackage{bm}
\usepackage{epsfig}
\usepackage{latexsym}
\usepackage{nameref,hyperref}
\usepackage{pifont}
\usepackage[utf8]{inputenc}
\usepackage{subcaption}
\usepackage{csquotes}
\usepackage{enumitem}
\usepackage{bm}
\usepackage{float}
\usepackage{xcolor}
\usepackage{multirow}
\usepackage{amsmath}
\usepackage{amssymb}
\usepackage[utf8]{inputenc}
\usepackage[T1]{fontenc}
\usepackage{lmodern} 
\usepackage{booktabs}

\DeclareUnicodeCharacter{0229}{\k{e}}

\begin{document}
\title{A Novel Mechanism of Ordering in a Coupled Driven System: Vacancy Induced Phase Separation}
\author{Chandradip Khamrai and Sakuntala Chatterjee}
\affiliation{Department of Physics of Complex Systems, S. N. Bose National Centre for Basic Sciences, Block JD, Sector 3, Salt Lake, Kolkata 700106, India.}

\begin{abstract}

We study a coupled driven system where two different species of particles, along with some vacancies or holes, move on a landscape whose shape fluctuates with time. The movement of the particles is guided by the local shape of the landscape, and this shape is also affected by the presence of different particle species. The nature of this coupling plays a crucial role in formation of long range order in the system. When a particle species push the landscape in the same (opposite) direction of its own motion, it is called an aligned (a reverse) bias. Aligned bias promotes ordering while reverse bias destroys it. In absence of vacancies, the system reduces to previously studied LH model for which different kinds of ordered and disordered phases were observed. This phases could be explained as a competition or cooperation between aligned bias and reverse bias coming from different particle species. This interplay is expected to remain unaffected even when vacancies are present since vacancies do not impart any kind of bias on the landscape. However, we find presence of vacancies effectively weakens the reverse bias and this significantly changes the outcome of the competition between the two bias types. As a result novel ordered phases emerge which were not seen before. We analytically calculate the new phase boundaries within mean field approximation. We show even when aligned bias is weaker than reverse bias, it is possible to find long range order in the system. We discover two new phases where particle species showing weak aligned bias phase separate and the other species with strong reverse bias stays mixed with the vacancies. We call these phases finite current with partial phase separation (FPPS) and vacancy induced phase separation (VIPS). The landscape beneath the phase separated species takes the form of a macroscopic hill or valley in FPPS phase. But in VIPS phase it has the shape like a plateau whose height scales as square root of system size. The landscape in the remaining part of the system is disordered in both these phases.

\end{abstract} 
 
\maketitle

\section{Introduction} \label{sec:intro}

Two or more nonequilibrium systems whose time evolutions are coupled to each other can show a wide range of interesting behavior depending on the nature of their coupling  \cite{singleactiveslider, bisht2019interface, singha2018time, singha2018clustering, sediments, ramaswamy2002phase, passiveslider1, passiveslider2, dolai2020universal, krapivsky2015tagged, krapivsky2015dynamical, malakar2020steady}. Sometimes such systems can show formation of ordered structures \cite{yu2022perpendicular, de2020flow, li2012formation, drossel2000phase, das2016phase, goswami2008nanoclusters, van2010hotspots, protein, veksler2007phase, gov2006dynamics, veksler2009calcium, kabaso2011theoretical, peleg2011propagating, fovsnarivc2019theoretical, yu2011early,  suarez2023reconstitution, litschel2024membrane, cagnetta2018active, chatterjee2007dynamics}, sometimes the coupling can give rise to metastable states \cite{Chaudhuri2006, kwak2004driven, yu2022perpendicular}, or sometimes the coupling is such that the system becomes disordered and showcases a variety of unconventional dynamical universality classes \cite{fibonacci, nonlinear, popkovuniversality, chakrabortyUniversality, popkov2014superdiffusive}. For example, in cell membrane the movement of certain proteins, called GPI-anchored proteins, is coupled to the dynamics of actomyosin network in cell cytoskeleton \cite{goswami2008nanoclusters, van2010hotspots, protein}, which results in clustering of the GPI-anchored protein molecules, as well as membrane structures like filopodia, lamellipodia, or microvilli \cite{suarez2023reconstitution, litschel2024membrane, cagnetta2018active}. Sometimes a nonequilibrium coupling can give rise to highly interesting dynamical effects as found in transport through a single-walled carbon nanotubes in aqueous environment where oscillatory behavior and noise-optimized transport were reported \cite{lee2010coherence}. Sometimes the nonequilibrium coupling drives the system to a disordered state, where tools of linear and nonlinear hydrodynamics \cite{popkov2004hydrodynamic, popkovexact, saito2021microscopic, miron2019derivation, chen2018exact, nonlinear, mendl2013dynamic, spohn2015nonlinear, prakash2025exact} can be used to study various phenomena like anomalous transport and mode-coupling phenomena.

A rather well-studied biological system which has received attention from physicists for the last few years is the coupled dynamics of cell membrane and membrane-bound proteins \cite{peleg2011propagating, veksler2007phase, gov2006dynamics, veksler2009calcium, kabaso2011theoretical, fovsnarivc2019theoretical, legg2007n, bj2004bar}. These proteins show preference towards particular type of local curvature. Some of them tend to localize in regions where the cell membrane is convex while others prefer concave parts of the membrane. Not only the proteins are guided by the shape of the membrane, they also modify the membrane shape by inducing local curvature at their positions. In absence of these proteins the membrane remains flat. Motivated by this, we consider a simple model that captures the effect of competing influences of different moving species on a membrane or landscape whose shape fluctuates with time. The model we study here does not aim to describe the specific biological system mentioned above and does not take into account all intricate details present in such systems. Instead we study a general model whose conclusions remain applicable or open for testing in a wider class of systems. In our model we have two different species of hardcore particles, named as `heavy' ($H$) and `light' ($L$), that move stochastically on a landscape whose height fluctuates in space and time. Heavy particles tend to slide downward along the local height gradient of the landscape, while light particles show a tendency to move upward along the gradient. Just like the landscape height profile affects the movement of the particles, the particles in turn also affect the landscape dynamics by pushing the local landscape height downward or by pulling it upward. The behavior of the system crucially depends on whether a particle species biases the landscape height in the same direction or opposite direction of its own motion. For example, $H$ particles prefer to move downward, and if they also push the landscape  downward, then it is called an aligned bias. In the case where $H$ particles pull the landscape upward, it is called a reverse bias. Similarly, $L$ particles pulling the landscape upward (downward) are termed aligned (reverse) bias. Apart from $L$ and $H$ particles there are also vacancies on the landscape, which are often called `holes' and they do not bias the landscape in any direction. We name the model light-heavy-vacancy model or LHV model in short.

In the absence of any vacancies when the entire landscape is covered by $L$ and $H$ particles, we have LH model which was introduced in \cite{chakrabortylarge}. Depending on which particle species applies aligned bias and which one applies reverse bias on the landscape, the LH model shows an interesting phase diagram comprising of different kinds of ordered and disordered phases \cite{static, dynamic, chakrabortyUniversality, khamrai2024effect, prakash2025exact, chakrabortyUniversality, lightheavy}. A general criterion that governs these phases can be stated as: aligned bias promotes ordering, while reverse bias destroys it. Based on this criterion, all phases of LH model can be described as competition or cooperation between these two types of bias. Our main motivation behind the present work is to investigate whether the same criterion remains valid even when LH model is generalized by allowing some vacancies in the system. The vacancies themselves do not impart any bias, and therefore is not expected to affect the interplay between aligned and reverse bias. Therefore, one might expect the above criterion to remain unchanged and the phases of LHV model to be same as those of LH model. However, our study shows presence of vacancies changes the phase diagram completely and new kinds of ordered phases emerge, which were not seen before.

More specifically, we show that the presence of vacancies effectively weakens the reverse bias and this significantly changes the outcome of the competition between the two bias types.   Earlier in LH model it was not possible to observe ordering if aligned bias was weaker than reverse bias \cite{chakrabortylarge}. But presence of vacancies makes it possible in the LHV model. The species that shows weaker aligned bias can still phase separate in a macroscopic hill or valley giving rise to macroscopic ordered structure in the landscape, while the species showing stronger reverse bias remains mixed with the vacancies and occupies a disordered segment of the landscape. Importantly, presence of the vacancies reduces the effective bias on the disordered segment. This makes it possible for the ordered and disordered parts of the system to coexist. The reverse bias, although larger in magnitude, is thus unable to destroy the ordering produced by the aligned bias because vacancies bring down the effective strength of the reverse bias. This new phase is called FPPS (finite current with partial phase separation).

When the reverse bias becomes much stronger than the aligned bias, presence of vacancies is not enough to preserve the order-disorder coexistence mentioned above. A new phase emerges at this point. We call it vacancy induced phase separation or VIPS in short. In this phase also there is an ordered segment and a disordered segment. But the ordered segment is different from FPPS phase. Here, the particle species which exert weak aligned bias still phase separate but do not form a pure phase. A small fraction of the other particle species with strong reverse bias is also mixed with it. The landscape underneath does not form a sharp hill or valley as before. Instead it takes the shape of a plateau, {\sl i.e.} a hill with a flat top (or a valley with a flat bottom, depending on the bias direction). This is a very new kind of ordered state not seen before. The rest of the system remains disordered as  seen for FPPS. We derive the condition for VIPS analytically within mean field approximation using flux balance condition in steady state. \textcolor{black}{We also explain the underlying physical mechanism which gives rise to the novel phase ordering by mapping one portion of the landscape to one of the paradigmatic model of nonequilibrium statistical physics, partially asymmetric exclusion process with open boundary.}

In the next section we describe the model. Then in Sec. \ref{sec:LHmodelRev} we give a brief overview of the LH model phase diagram. In Sec. \ref{sec:steadystate} we present phase diagram for the new LHV model. The newly discovered VIPS phase is discussed in detail in Sec. \ref{sec:VIPSOrdering}. Sec. \ref{sec:Conclu} contains some concluding remarks.

\section{Model description} \label{sec:model}

\begin{figure}[H]
\centering
\includegraphics[scale=0.75]{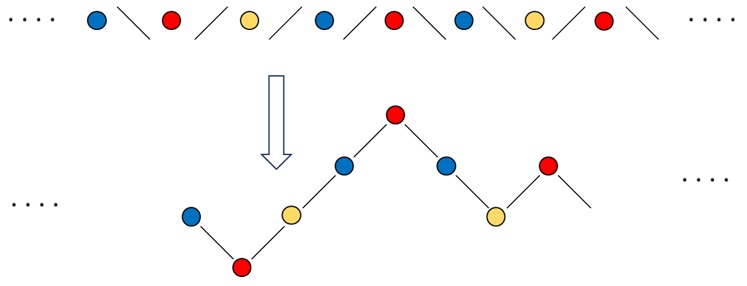}
\caption{Mapping onto a height model. Heavy, light particles and holes are respectively shown as red, yellow and blue circles. Upslope and downslope bonds are depicted by $\slash$ and $\backslash$ respectively.} 
\label{fig:hconf}
\end{figure}

Our model is defined on a one-dimensional lattice whose sites are either occupied by a `light' $(L)$ particle or a `heavy' $(H)$ particle or the site can remain vacant $(V)$. A vacant site is also sometimes called a hole. We denote the site occupancy by a variable $\eta_{j}$ which takes the value $+1$, $-1$ or $0$ if the site $j$ is occupied by an $H$ particle, $L$ particle or a hole, respectively. Particles are hardcore, which means no two particles can occupy the same site. The bonds of the lattice between each neighboring pair of sites can have two possible orientations, $\pm \pi /4$, and we use a variable $\tau_{j + 1/2}$ to denote it. A bond with orientation $\pi /4$ ($-\pi /4$) is called an upslope (a downslope) bond and is represented by $\tau_{j + 1/2} = 1$ ($-1$). We also define a height field $h_i = \sum_{j=1}^{i-1}  \tau_{j + 1/2}$, and assign a height to each site $i$. In a pictorial depiction of a lattice configuration, we use the symbol $\slash$ for an upslope bond and $\backslash$ for a downslope bond. To show the site occupancies we use a red circle for an $H$ particle, a yellow circle for an $L$ particle, while an empty site is represented by a blue circle. In Fig. \ref{fig:hconf}, we show a typical configuration with different species of particles on a surface (or landscape) with spatially varying height profile.  

\begin{figure}[H]
\centering
\includegraphics[scale=0.6]{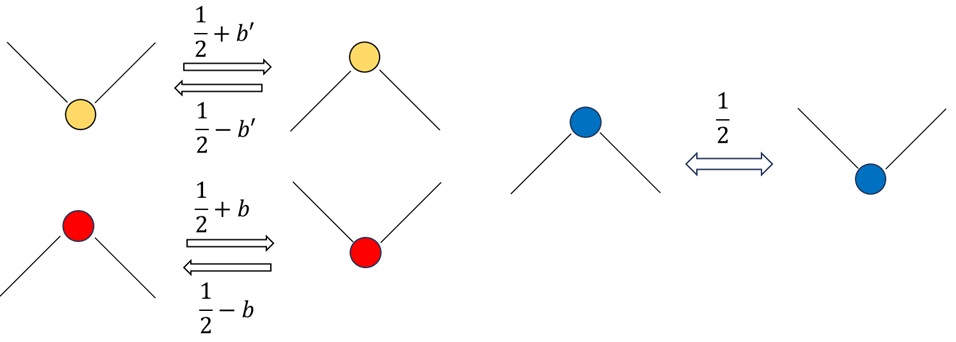}
\caption{Transition rules between the local hills and valleys of the landscape occupied by either $H$, $L$ particles or holes. Here $-\frac{1}{2} \leq b, b' \leq \frac{1}{2}$.} \label{fig:smoves}
\end{figure}

The transition rules in the model are such that the movement of the particles is affected by the surface, and the dynamics of the surface is affected by the local particle occupancies. The surface evolves by interchanging the orientation of the neighboring bonds. An upslope bond followed by a downslope bond is called a local `hill' while the reverse order is known as a local `valley'. The transition probabilities between hill and valley depend on the occupancy of the intervening site. As shown in Fig. \ref{fig:smoves}, a hill occupied by an $H$ particle can switch to a valley with probability $(\frac{1}{2}+b)$, while the reverse transition happens with probability $(\frac{1}{2}-b)$. Clearly, for $b > 0$ hill to valley transition occurs with a greater rate and  the reverse transition becomes slower. The $H$ particles are then said to push down the surface. On the other hand, $b < 0$  results in $H$ particles pulling up the surface. Similarly, a valley occupied by an $L$ particle can flip to a hill with a probability $(\frac{1}{2}+b')$ and the reverse transition happens with a probability $(\frac{1}{2}-b')$. A positive (negative) $b'$ means $L$ particles pull up (push down) the surface. If the intervening site is vacant, both forward and reverse transitions happen with the same probability $\frac{1}{2}$. We show all these moves in Fig. \ref{fig:smoves}. Both $b$ and $b'$ are bounded between $[-\frac{1}{2},\frac{1}{2}]$. 

\begin{figure}[H]
\centering
\includegraphics[scale=0.5]{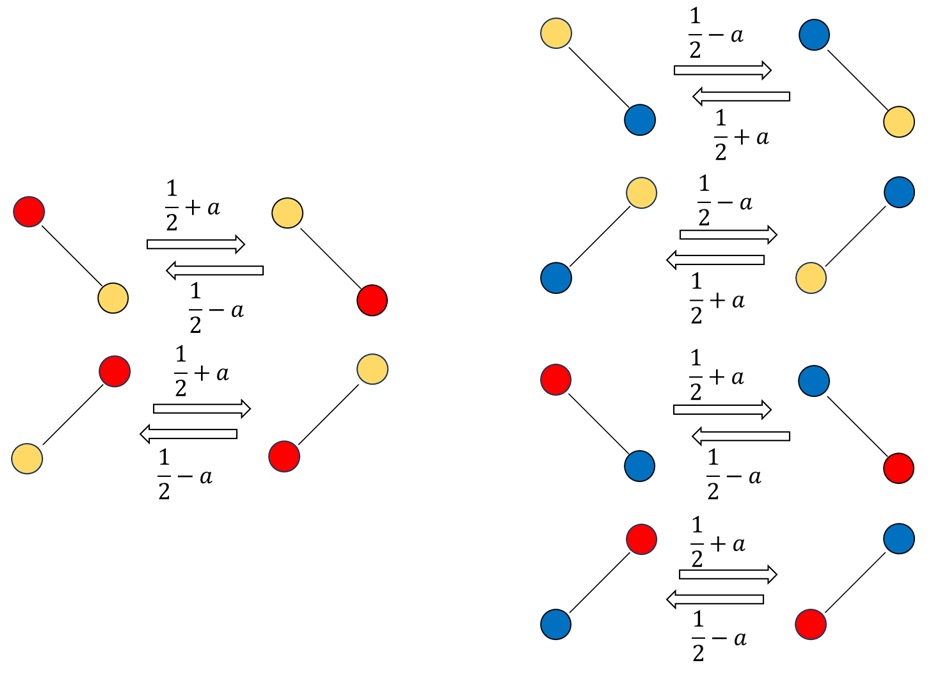}
\caption{Transition rules for the particles where $0 < a \leq \frac{1}{2}$.} \label{fig:ParUpdate}
\end{figure}

The particles move from one site to a neighboring site by sliding along the in-between bond. $H$ particles prefer to slide downward and $L$ particles prefer to slide upward. This preference decides the rate at which an $H$ ($L$) particle exchanges its position to a neighboring $L$ ($H$) particle or hole. For example, if as a result of this exchange the $H$ particles move to a site with lower height, then it happens with a probability $(\frac{1}{2}+a)$, while the reverse transition happens with a probability $(\frac{1}{2}-a)$, with $0 < a \leq \frac{1}{2}$. We show the moves in Fig. \ref{fig:ParUpdate}.

In the absence of any vacant sites, a closely related model was introduced in \cite{chakrabortylarge} as LH (light-heavy) model and was extensively studied in \cite{dynamic, static, chakrabortyUniversality, khamrai2024effect}. Here, we introduce vacant sites and call it LHV (light-heavy-vacancy) model. We use periodic boundary condition on a lattice of $N$ sites, with $N/2$ number of upslope bonds and remaining downslope bonds. \ We represent the total number of $H$ and $L$ particles respectively by $N_H$ and $N_L$; so the number of vacant sites becomes $(N - N_H - N_L)$. Densities of $H$ and $L$ particles are denoted by $\rho_{H} = N_{H}/N$ and $\rho_{L} = N_{L}/N$, respectively. Most of our data presented here are for $\rho_H = \rho_L = 1/3$ unless mentioned other wise. However, all our conclusions for LHV model remain valid for any finite densities of $H$, $L$ particles and vacancies. All simulation data presented in this paper have been averaged over at least $10^6$ histories.

For $b > 0$, an $H$ particle which prefers to slide downward, also pushes the surface downward. This is called an aligned bias. For $b < 0$ the $H$ particles pull the surface upward, which is opposite to its own preferred direction of motion. This is called a reverse bias. Similarly, for $L$ particles $b' > 0$ corresponds to aligned bias and $b' < 0$ means reverse bias. We will show below how phase diagram of the model sensitively depends on the nature of the bias. 

\section{Overview of different phases in LH model} \label{sec:LHmodelRev}

Without any vacant sites, the model reduces to the LH model, which was introduced in \cite{chakrabortylarge}. By changing the two parameters $b$ and $b'$ a phase diagram for LH model was obtained in an earlier study \cite{chakrabortylarge}  which consists of different types of ordered and disordered phases \cite{static, dynamic}. In this section, we give a brief overview of these phases. 

\begin{figure}[H]
\centering
\includegraphics[scale=0.8]{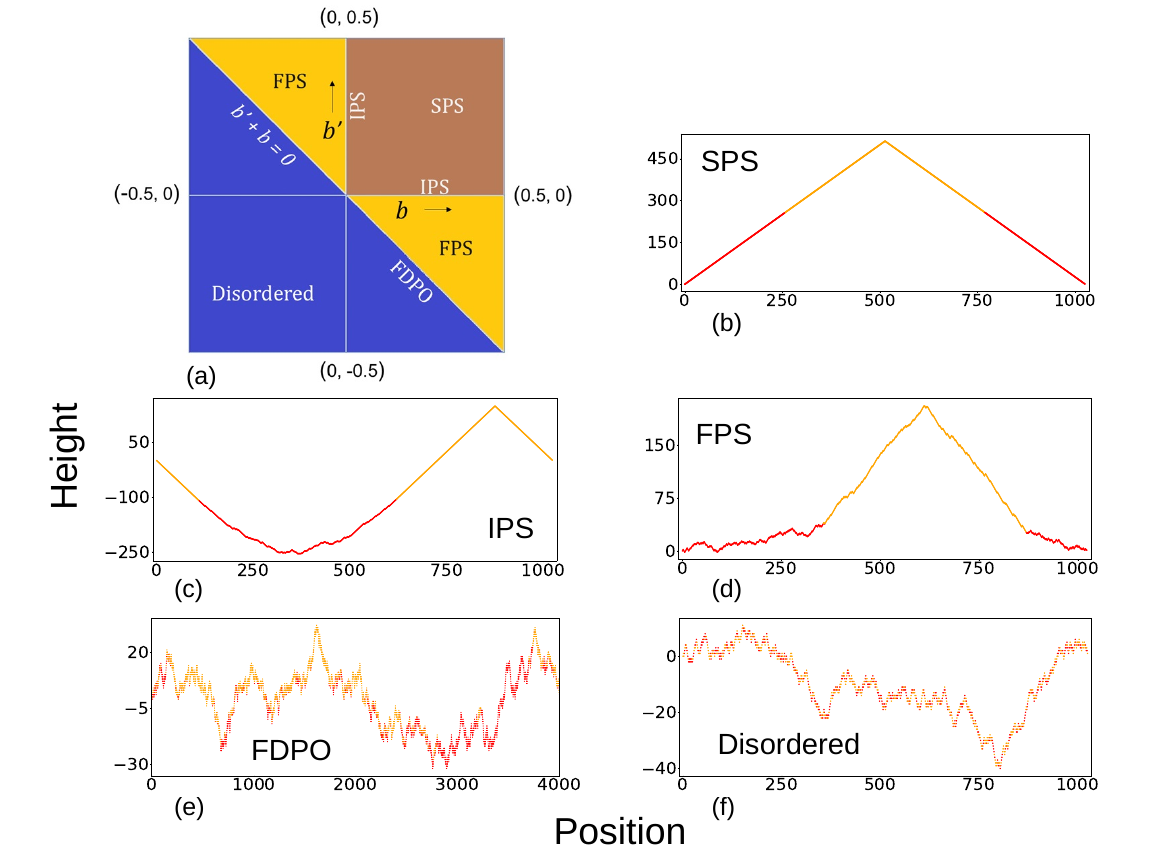}
\caption{LH model phase diagram and representative configuration in each phase. The colours red, and yellow represents $H$, and $L$ particles, respectively. The values of $b,b'$ are as follows.(b) SPS, $b = b' = 0.4$; (c) IPS, $b' = 0.4$, $b = 0$; (d) FPS, $b' = 0.4$, $b = -0.1$; (e) FDPO, $b' = 0.4$, $b = -0.4$; (f) disordered phase, $b = b' = -0.4$. Here, $N = 1026$, $\rho_{H} = \rho_{L} = 1/2$ and $a = 0.4$. This phase diagram is valid for any finite density of heavy and light particle at thermodynamic limit, $N \to \infty$.}
\label{fig:lhconfig}
\end{figure}

\textbf{Strong Phase Separation (SPS)}: This phase is obtained when both $b$ and $b'$ are positive, as shown in Fig. \ref{fig:lhconfig}(a). In this case, $H$ particles push the surface down, while $L$ particles pull it upward. In other words, the $H$ particles which prefer to occupy the valleys, also make the valleys stable. Similarly, the $L$ particles stabilize the hills, which are also their preferred locations. This kind of positive feedback gives rise to maximum possible order in the system. The upslope bonds completely phase separate from the downslope bonds to give rise to a single large valley and a single large hill in the system. All $H$ particles form a single cluster which occupies the valley, while a single cluster of $L$ particles is present in the hill. In Fig. \ref{fig:lhconfig}(b) we show a typical configuration. 

\textbf{Infinitesimal current with Phase Separation (IPS)}: This phase is obtained when one particle species does not push or pull the surface in any direction, while the other applies an aligned bias. In this case both the particle species phase separate completely, while the landscape shows partial order. For example, when $b=0, b' > 0$ the surface beneath the $L$-cluster orders itself to form a macroscopic hill, but the surface occupied by the $H$ particles does not show perfect order. Instead of phase separation between upslope and downslope bonds, this segment of the landscape shows a spatial gradient of these bonds which scales as $1/N$. The shape of the surface is parabolic in this part. The whole surface moves upward with a velocity $\sim 1/N$. In Fig. \ref{fig:lhconfig}(c) we show a representative configuration. IPS phase is also seen when the $H$ particles push the surface down while the $L$ particles remain neutral, {\sl i.e.} $b > 0, b' = 0$. Static and dynamic properties of IPS phase were studied earlier in \cite{chakrabortylarge, static, dynamic}.

\textbf{Finite current with Phase Separation (FPS)}: This phase is obtained when both particle species try to move the surface in the same direction, but at different rates. For $b' > -b > 0$ both $H$ and $L$ particles pull the surface upward but $L$ particles do so at a higher rate. Similarly, for $b > -b' > 0$ both particle species push the surface downward but $H$ particles push at a faster rate. This phase again shows complete phase separation between $H$ and $L$ particles, but the landscape beneath one species shows ordering, while the remaining landscape stays disordered. For $b' > -b > 0$, the $L$-cluster occupies a large hill, while the surface beneath the $H$-cluster has equal densities of upslope and downslope bonds, with roughly equal average height everywhere. The entire surface moves upward with a finite velocity in this case. We show a representative configuration in Fig. \ref{fig:lhconfig}(d). Similarly, for $b > -b' > 0$ the $H$-cluster is present in a macroscopic valley while the $L$-cluster is found on a disordered surface segment. FPS phase was studied in detail in  \cite{static, dynamic}.

\textbf{Fluctuation Dominated Phase Ordering (FDPO)}: For $b + b' = 0$, both particle species affect the surface in an identical manner, {\sl i.e.} with the same rate they either push the surface down or pull it upward. So irrespective of whether an $H$ or $L$ is present, the hill-valley transitions happen with the same rate everywhere on the surface. In this limit the coupling between the particles and the surface becomes effectively one-way. This is the limit of passive scalar advection. In steady state the surface remains disordered and the $H$ and $L$ particles do not show complete phase separation as seen in SPS, IPS or FPS phases. Instead the particles show an unconventional kind of ordering where large clusters and strong fluctuations coexist. In \cite{dasparticles, chatterjee2006dynamics, das2001fluctuation} this phase was studied in detail.

\textbf{Disordered phase}: In the parameter space $b + b' < 0$, disordered phase emerges under the following conditions: (i) both $H$ and $L$ particles pull the surface upward, with $H$ pulling at a higher rate; (ii) both species push the surface downward, but $L$ particles do so more strongly; or (iii) $H$ particles pull the surface upward while $L$ particles push it downward. In all these cases both the surface and the particles show no long-range order. In \cite{chakrabortyUniversality} the dynamical properties of this phase were studied using tools of nonequilibrium fluctuating hydrodynamics.

To summarize, depending on the sign of $b$ and $b'$, the preferred direction of movement of a particle species (which is upward for $L$ and downward for $H$) may align with, or may be opposite to, the direction in which they are biasing the surface. While aligned bias promotes ordering, reverse bias destroys it. The phases SPS, IPS, FPS have one thing in common: there is at least one particle species which applies aligned bias. This ensures those particles will form a single cluster and the surface beneath that cluster will also show long range-order. Clearly, the second species of particles also forms a single cluster then, irrespective of how they bias the surface. But the nature of ordering present in the surface beneath this second cluster depends on this bias. When this second bias is also aligned we have SPS and the surface is perfectly ordered. For zero bias, we have IPS. For reverse bias, whose magnitude is smaller than the aligned bias coming from the first species, we have FPS when the surface beneath the second species is completely disordered. When the reverse bias from the second species is larger than the aligned bias from the first species, no long-range order is possible for particles and surface and we have disordered phase. In the next section, we see how the presence of vacancies modifies these criteria.

\section{Phase diagram for LHV model} \label{sec:steadystate}

\begin{figure}[H]
\centering
\includegraphics[scale=0.8]{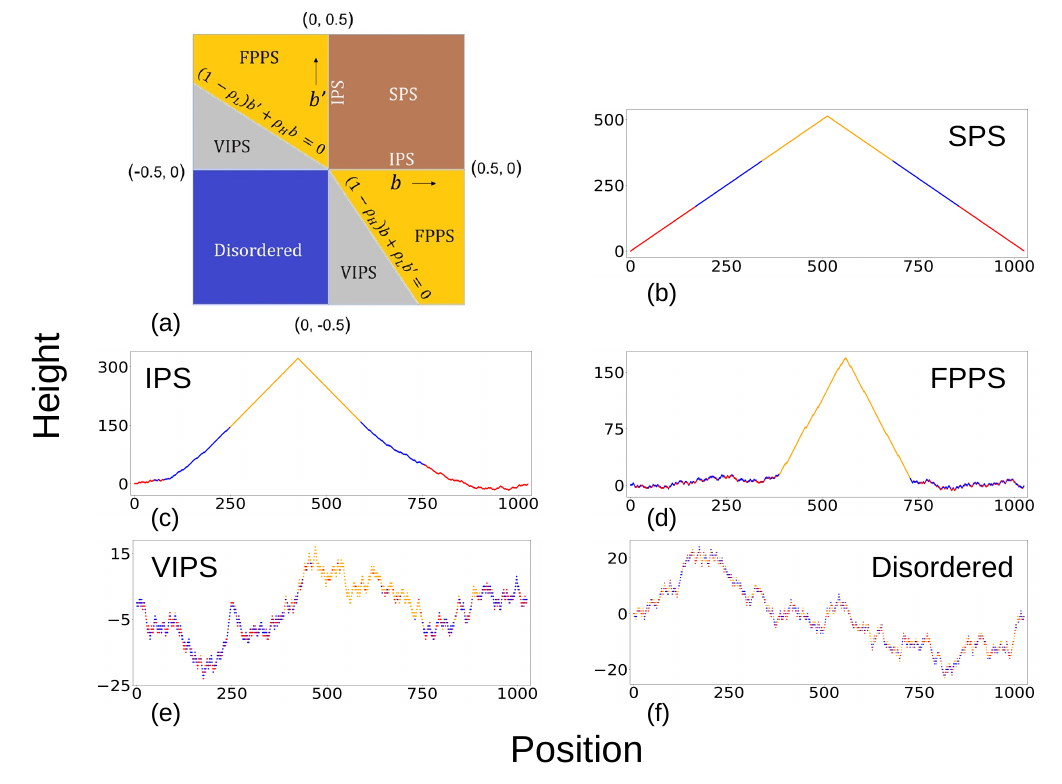}
\caption{Phase diagram for LHV model and representative configurations in each phase.  The colours red, yellow and blue represent $H$, $L$ particle and hole, respectively. The values of $b,b'$ are as follows: (b) SPS, $b = b' = 0.4$; (c) IPS, $b' = 0.4$, $b = 0$; (d) FPPS, $b' = 0.4$, $b = -0.1$; (e) VIPS, $b' = 0.18$, $b = -0.4$; (f) disordered phase, $b = b' = -0.4$. All representative configurations are shown for $N = 1026$, $\rho_{H} = \rho_{L} = 1/3$ and $a = 0.4$. The phase diagram  is valid for any finite density of different particle species and vacancies at the thermodynamic limit, $N \to \infty$.} 
\label{fig:phasev}
\end{figure}

In presence of vacancies, the ordered and disordered phases of LHV model are shown in Fig.\ref{fig:phasev}. When both $H$ and $L$ particles impart aligned bias on the surface, we have SPS (strong phase separation) phase and when one species applies zero bias, we have IPS (infinitesimal current with phase separation) phase. From representative configurations shown in Figs. \ref{fig:phasev}(b) and (c), it is seen that in these two phases, different species of particles show complete phase separation and the vacant sites or holes place themselves in between the $L$ and $H$ clusters. These two phases are analogous to those observed in the LH model.
  
However, new phases emerge when one species applies aligned bias and the other one applies reverse bias. For example, when $b' > 0, b < 0$, {\sl i.e.} both $L$ and $H$ particles pull the surface upward, which means bias from $L$ is aligned while that from $H$ is opposite. In this case, we find that for small magnitudes of $b$, all $L$ particles phase separate into a single cluster and occupy a macroscopic hill (see Appendix \ref{sec:FPPSPhase}). The $H$ particles and holes are mixed together and occupy the rest of the surface which remains disordered. Although the ordered part of the system is closely similar to that in FPS phase in LH model, it coexists with a disordered part, where $H$ particles displace the holes to occupy the valleys and also make the valleys unstable. No long-range order can form in this part. $H$ particles which apply reverse (upward) bias on the surface are mixed with holes which stay neutral. The dynamical rules of this part of the system are similar to the original LH model with  $b<0, b'=0$, {\sl i.e.} the line separating the second and third quadrant in Fig.\ref{fig:lhconfig}(a), which belongs to the disordered phase of the LH model. Moreover, since both $L$ and $H$ pull the surface upward, the whole surface moves upward with a finite velocity. We describe this new phase as `finite current with partial phase separation' or FPPS in short. We use the phrase `partial' phase separation as only one particle species is completely phase separated. This phase can be observed as long as the magnitude of $b$ remains small enough. When $b$ becomes too large, $H$ particles pull the surface upward at a much larger rate than $L$, then the macroscopic hill beneath the $L$ cluster can not be sustained anymore. This marks the boundary of FPPS phase. Below we analytically derive the equation of this phase boundary within mean field approximation.

We start with the large hill in the ordered part of the FPPS phase. Such a hill consists of a macroscopic segment of surface with majority upslope bonds mixed with few downslope bonds, followed by another segment of roughly equal length with majority downslope bonds along with few upslope bonds. Let $\lambda$ denote the fraction of minority bond type present in each of these segments. Clearly, we need $0 < \lambda < 1/2$ for the macroscopic hill to exist. This hill is occupied by the $L$ cluster in the FPPS phase. If $P(/L\backslash)$ denotes the probability to find a microscopic hill occupied by an $L$ particle, then $(0.5-b') P(/L\backslash)$ is the flux of probability associated with hill to valley transition. Similarly, $(0.5+b') P(\backslash L /)$ is the probability flux associated with the reverse transition. The difference between these two fluxes determines the upward velocity of this segment. We write it as
\begin{equation}
J_1 =  \Bigl(\frac{1}{2}+b'\Bigr) P(\backslash L /) - \Bigl(\frac{1}{2}-b'\Bigr) P(/L\backslash)
\end{equation}
Using mean field approximation we factorize these probabilities 
\begin{equation}
J_1 =2 b' \lambda (1-\lambda) 
\end{equation}
Similarly, for the disordered segment of the surface, where $H$ particles are present along with vacant sites, the upward velocity is 
\begin{equation}
J_2 = \biggl[\Bigl(\frac{1}{2}-b\Bigr)P(\backslash H /) - \Bigl(\frac{1}{2}+b\Bigr)P(/H \backslash)\biggr] = -\frac{b \rho_H}{2(1-\rho_L)}
\end{equation} 
since the vacant sites not impart any bias on the surface. Note that the current here depends on the densities of the particles. In the limit of zero density of vacancies, when we have LH model, $J_2 = -b/2$ gives the current in FPS phase which does not depend on densities. Also, in this disordered part it is equally likely to find an upslope or downslope bond. In steady state we must have $J_1 = J_2$ which gives 
\begin{equation}
2 b' \lambda (1-\lambda) = -\frac{b \rho_H}{2(1-\rho_L)}.
\end{equation}
This equation can be solved for $\lambda$
\begin{equation}
\lambda = \frac{1}{2}\Biggl[1 - \Bigl(1 + \frac{\rho_{H}}{1-\rho_L}\frac{b}{b'}\Bigr)^{1/2}\Biggr]
\label{eq:minden2}
\end{equation} 
The large hill in FPPS exists as long as $\lambda < 1/2$, {\sl i.e.} $(1-\rho_L) b' + \rho_H b > 0$. Thus the boundary of the FPPS phase is marked by $\lambda = 1/2$ or 
\begin{equation}
(1-\rho_L) b' + \rho_H b = 0.
\end{equation}
Note that in absence of any vacancies as in LH model, $(1-\rho_L) = \rho_H$ and we get $b+b'=0$ which is the equation of FDPO line shown in Fig.\ref{fig:lhconfig}.

Interestingly, by choosing $\rho_H$ and $\rho_L$ suitably, we can have FPPS phase even when $H$ particles pull the surface upward faster than $L$. This possibility was ruled out in the original LH model, where $b+b'<0$ always yielded disordered phase. In the LHV model vacant sites are present which do not bias the surface. In the FPPS phase these vacancies are mixed with the $H$ particles which bring down the effective upward velocity of this portion of the landscape. This allows the $L$ particles to phase separate in a macroscopic hill, even though their upward pull is smaller than $H$.

However, as $(1-\rho_L) b' + \rho_H b$ becomes negative, then $H$ particles pull the surface upward so much faster than $L$, that even though they are mixed with holes, the disordered segment of the landscape moves upward at a higher rate and catches up with the macroscopic hill containing the $L$ cluster. The macroscopic hill then becomes unstable. A small fraction of the $H$ particles roll into the $L$ dominated region to ensure that in steady state all parts of the surface move with the same velocity. The $L$ particles still phase separate but not in a pure cluster, a small but finite fraction of $H$ particles is always present in it, while the remaining $H$ particles are mixed with the holes as before in a disordered landscape. In Fig.\ref{fig:phasev}(e) we show a representative configuration. Although from this snap shot the landscape looks disordered everywhere, we show in the next section that the average height profile beneath the $L$ dominated region is not flat. In fact it has the shape of an elevated plateau, {\sl i.e.} a hill with flat top. The height of the plateau scales as  $\sqrt{N}$, in sharp contrast with the macroscopic hill in FPPS phase whose height was linear in $N$. This novel phase was not known earlier in LH model and we call it vacancy induced phase separation or VIPS in short. In Fig.\ref{fig:phasev}(a) we mark this phase in the phase diagram.

We have discussed above the specific example when both species of particles pull the surface upward. Similar results are obtained when both species push the surface downward, $b >0, b'<0$. It is easy to argue that in this case the $H$ particles phase separate in a macroscopic valley in the FPPS phase, and $L$ particles are mixed with holes in the disordered region of the surface. The phase boundary between FPPS and VIPS phase is  given by $(1-\rho_H) b + \rho_L b' =0$. In the next section, we discuss the VIPS phase in more details. Finally, when $b'$ and $b$ both are negative, both species of particles exert reverse bias on the surface. In this case no long range order is present for particles or surface, and we have a disordered phase.

\section{Ordering in VIPS phase} \label{sec:VIPSOrdering}

\begin{figure}[H]
\centering
\includegraphics[scale=1]{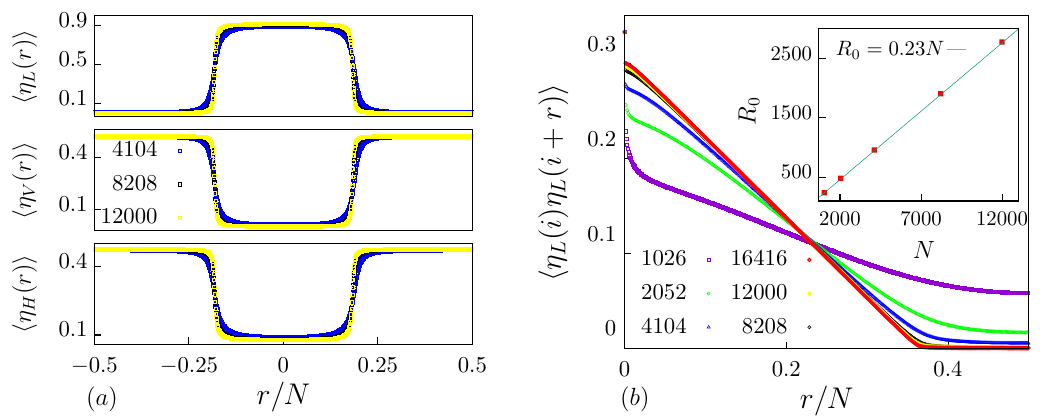}
\caption{(a) Site occupancy as a function of scaled distance $r/N$ from the center of mass of $L$ particles. Top, middle and bottom panels are for $L$ particles, vacancies and $H$ particles, respectively. (b) Two-point density correlation function for $L$ particles show scaling collapse for large $N$ when plotted against $r/N$, but for small $N$ scaling collapse of the full curve is not observed. The distance $R_0$ at which correlation function equals $\rho_L^2$ scales with $N$ (inset). Here we use $a = 0.4$, $b' = 0.18$ and $b = -0.4$. The densities of the particle species are $\rho_{H} = \rho_{L} = 1/3$.} \label{fig:ParOcc}
\end{figure}

To study the nature of ordering present in VIPS phase, we first look at the steady state density profile of each particle species. In Fig. \ref{fig:ParOcc}(a) we plot these densities as a function of the distance from the center of mass of the $L$ clusters. Unless mentioned otherwise in this section, we measure all distances from this point. Our data in Fig. \ref{fig:ParOcc} are for the case when $\rho_H = \rho_L = 1/3$. We monitor the variation of densities with scaled distance $r/N$ and our data show that for $|r| < N/6$ the density of $L$ particles remains close to $1$. This means that most $L$ particles are confined in this region. We also observe that the density of $H$ particles takes a small but finite value in this zone. So the $L$ cluster is not completely pure but a small number of $H$ particles is mixed in it. We estimate this fraction analytically using mean field approximation below. For $|r| > N/6$ the density of $L$ particles drops rapidly to almost zero and that of $H$ particles picks up to $1/2$. For $|r| > N/6$ we have the disordered segment where $H$ particles and holes are randomly mixed together. 

In Fig. \ref{fig:ParOcc}(b) we plot two-point density correlation function of $L$ particles with scaled distance $r/N$ for different system sizes $N$. For small $r$ the correlation function shows a sharp decay but this behavior does not extend beyond a finite range of $r$ even when $N$ increases. Since we are interested in the large $N$ limit, this small $r$ behavior is not so important for us. Indeed our data in Fig. \ref{fig:ParOcc}(b) show that when plotted against $r/N$, this range is hardly visible as $N$ becomes large. For large $N$ we find a scaling collapse of the entire curve, but for smaller $N$ values, there is no scaling collapse. In Appendix \ref{sec:HFraction} we discuss the finite size effects in more details. In the inset of Fig. \ref{fig:ParOcc}(b) we plot the distance $R_0$ at which the two point correlation function crosses the value $\rho_L ^2$, as a function of $N$ and find linear scaling. Presence of a macroscopic length scale in the two-point density correlation function clearly shows the phase separation of $L$ particles.

\begin{figure}[H]
\centering
\includegraphics[scale=1]{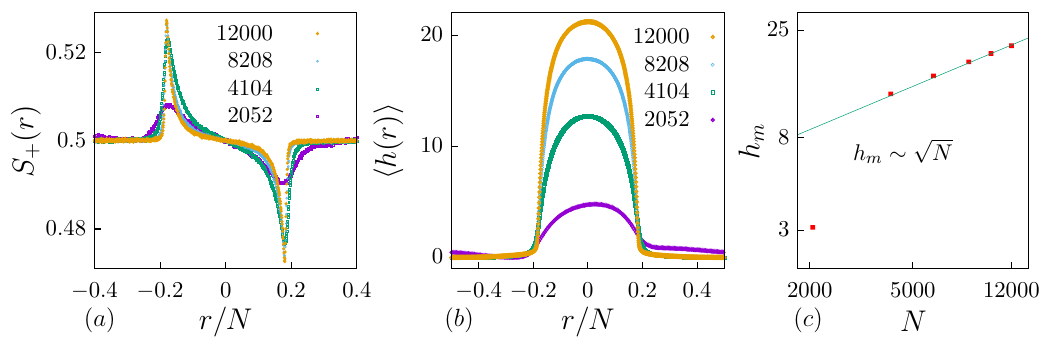}
\caption{(a) The probability of finding an up-slope bond $S^{+}(r)$ at a (scaled) distance $r/N$ from the center of mass of the $L$ particles for $N$. (b) The average height profile for different $N$. Inset: Height $h_m$ of the plateau scales as $\sqrt{N}$ when $N$ is large. For smaller $N$ there is significant deviation from this scaling, as seen for $N=2052$. Here we use $a = 0.4$, $b' = 0.18$ and $b = -0.4$. The densities of the particle species are $\rho_{H} = \rho_{L} = 1/3$.} \label{fig:rL}
\end{figure}

To understand the shape of the landscape in VIPS phase, we measure the probability $S_+(r)$ of finding an upslope bond at a (scaled) distance $r/N$ from the center of mass of $L$ particles. Our data in Fig. \ref{fig:rL}(a) show that exactly at the center $S_+(r=0) = 1/2$. For small but non-zero $r$, we find $S_+(r)$ varies from $1/2$ but the variation becomes slow as $N$ increases. We calculate the average height profile as $\langle h(r) \rangle = \sum_{x=-N/2}^r (2S_+(x) - 1)$ and show it in Fig. \ref{fig:rL}(b). We find plateau like structure, where the height at the top becomes almost flat. In Fig. \ref{fig:rL}(c) we plot the height $h_m$ at plateau top as a function of $N$ and find a $\sqrt{N}$ scaling for large $N$. This is distinctly different from the macroscopic hill in the FPPS phase which supports the $L$ cluster. 

\textcolor{black}{The VIPS phase constitutes a special case in which, due to the coupled dynamics of particles and the surface, the motion of slopes beneath the light-particle cluster effectively reduces to an one-dimensional partially asymmetric exclusion process\cite{sandow1994partially, blythe2007nonequilibrium} of up-slopes with open boundaries, operating in the maximal current phase. The remaining disordered landscape, containing heavy particles and vacancies, acts as a reservoir from which up-slopes enter the VIPS plateau through the left boundary and exit through the right boundary at equal rates. This mechanism provides the physical origin of the emergent VIPS phase. In the maximal current phase, it is well known that the bulk density remains close to $1/2$ and is essentially independent of the boundary rates, while boundary effects penetrate deep into the system through slowly decaying boundary layers—being higher than the bulk near the entrance and lower near the exit. The spatial variation of the up-slope occupation probability in fig.\ref{fig:rL} within the light-particle–occupied plateau region, $-\rho_LN/2 < r < \rho_LN/2$, exhibits precisely the same behavior.}

\textcolor{black}{We know that the in case of one-dimensional partially asymmetric exclusion process, in the maximal current phase, near the left boundary, the density is larger than $1/2$ and decreases towards the bulk as $\rho(x) \sim \frac{1}{2} + A_{\mathrm{L}}\,x^{-1/2}$, where $x$ denotes the distance from the entrance with $A_{\mathrm{L}}$ being non-universal amplitudes that depend on the boundary rates. Following this we can approximate the average up-slope occupation probability in the left half of the light particle occupied plateau as $S_+(x) \sim \frac{1}{2} + A_{\mathrm{L}}\,x^{-1/2}$, where $x$ denotes the distance from the left boundary. Applying this, average height of the center of the plateau can be calculated as}

\begin{equation}
\color{black}
h_m = \int_{0}^{\rho_LN/2} \bigl(2S_+(x) - 1\bigr) dx = C\sqrt{N}
\label{eq:BounDen3}
\end{equation}
\textcolor{black}{Where $C$ is some constant. So the $\sqrt{N}$ scaling of the plateau height is coming from the algebraic boundary relaxation of the up-slopes in the VIPS phase.} 

\begin{figure}[H]
\centering
\includegraphics[scale=1]{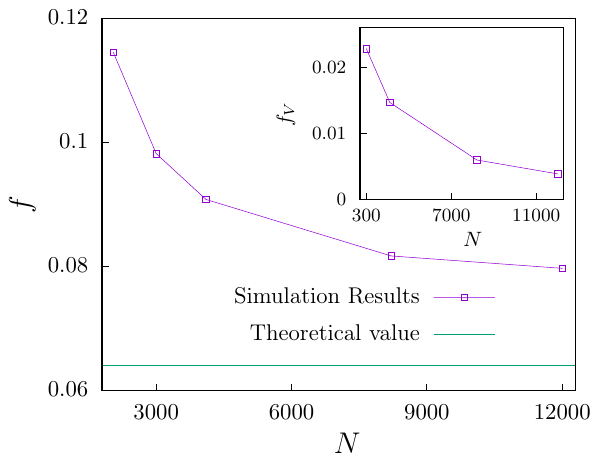}
\caption{Fraction of $H$ particles interspersed within the $L$ particle clusters decreases with $N$. The green line indicates the theoretical prediction from Eq.\ref{eq:CurrEq2}. Here we use $a = 0.4$, $b' = 0.18$ and $b = -0.4$. The inset shows the fraction of vacancies present in between the $L$ clusters which is negligibly small for large $N$. The densities of the particle species are $\rho_{H} = \rho_{L} = 1/3$.} \label{fig:Hfrac}
\end{figure}

From our plot of density profile in Fig. \ref{fig:ParOcc}(a) it is seen that in the region where $L$ particles phase separate, a small number of $H$ particles are also present. This ensures in steady state all parts of the surface move upward with the same average velocity. We can use this criterion to estimate the number of $H$ particles which are mixed with the $L$ particles. Out of a total number $N_H$ of $H$ particles, let a fraction $f$ be present in the $L$ dominated region. The upward velocity in this region can be written as 
\begin{equation}
\Bigl(\frac{1}{2}-b\Bigr)P(\backslash H /) - \Bigl(\frac{1}{2}+b\Bigr)P(/ H \backslash) - \Bigl(\frac{1}{2}-b'\Bigr)P(/L \backslash) + \Bigl(\frac{1}{2}+b'\Bigr)P(\backslash L /). 
\label{eq:topv}
\end{equation}
We assume in the plateau region it is equally likely to find an upslope or downslope bond. Then using mean-field approximation Eq. \ref{eq:topv} becomes
\begin{equation}
 \frac{(b'N_L - bfN_H)}{2(fN_H + N_L)}. \label{eq:topvmft}
\end{equation}
Similarly, in the disordered segment of the surface where $(1-f)N_H$ number of $H$ particles are present along with $N_V$ vacancies, the upward velocity can be written as 
\begin{equation}
\Bigl(\frac{1}{2}-b\Bigr)P(\backslash H /) - \Bigl(\frac{1}{2} + b\Bigr)P(/ H \backslash) = - \frac{b(1-f)N_H}{2[N_V + (1-f)N_H]} \label{eq:disvmft}
\end{equation}
Equating Eq. \ref{eq:topvmft} and r.h.s. of Eq. \ref{eq:disvmft} and dividing numerator and denominator by $N$, we get the following expression for $f$
\begin{equation}
f = \frac{\rho_L}{\rho_H}\biggl[\frac{\rho_Hb + (1 - \rho_L)b'}{\rho_Lb' + (1 - \rho_H)b}\biggr].
\label{eq:CurrEq2}
\end{equation}
In Fig. \ref{fig:Hfrac} we compare $f$ with our simulations. For smaller $N$ values we find significant deviations and for large $N$ the deviation is less. However, even for the largest possible $N$ we could access in simulations, there is still a mismatch between our mean field prediction and simulation results.  One possible reason for this could be breakdown of mean field approximation. However, even in the absence of a quantitative agreement, Eq. \ref{eq:CurrEq2} still provides useful insights. If we put $f=0$ in Eq. \ref{eq:CurrEq2} we get
\begin{equation}
\frac{\rho_H}{1 - \rho_L}b + b' = 0
\label{eq:CurrEq3}
\end{equation}
which is the equation for FPPS-VIPS phase boundary derived in the previous section. Exactly at this boundary $L$ phase is still pure, but as we go deeper inside the VIPS phase, the reverse bias from $H$ particles becomes so much stronger that a few $H$ particles must be mixed with the $L$ particles to ensure steady state. 

Fig. \ref{fig:ParOcc}(a) middle panel also shows a very small fraction of vacancies are also present in the $L$ cluster. In the inset of Fig. \ref{fig:Hfrac} we plot this fraction $f_V$ for different $N$ values. We find $f_V$ is negligible (one order of magnitude less) compared to $f$. Our plot also indicates $f_V$ decreases with $N$. For large enough $N$ presence of a few holes in the plateau region can therefore be neglected.

Note that both FPPS and VIPS phases contain a disordered segment where $H$ particles and holes are mixed. The fraction of $H$ particles present in the disordered sector is smaller in VIPS phase where $f > 0$. This implies for a given value of $b$, the surface current has a smaller magnitude in VIPS than in FPPS phase.

\section{Conclusions} \label{sec:Conclu}

In this paper, we explore different phases of LHV model where light, heavy particles along with some vacancies move on a fluctuating landscape. Previous work on a related model, known as LH model, had shown that the nature of ordering present in the system depends crucially on whether the bias applied by the particles on the landscape is aligned with or opposite to their own preferred direction. Aligned bias causes long range order and reverse bias disrupts any ordering. As a result, when one particle species shows aligned bias and the other species shows reverse bias, the resulting state is decided by which bias is stronger. When aligned bias wins over reverse bias, LH model shows FPS (finite current with phase separation) phase, which has long range order. But when reverse bias is stronger, LH model shows disordered phase. On the other hand, in the LHV model disordered phase is observed only when both particle species show reverse bias. As long as at least one species has aligned bias, LHV model shows long range order, even if the other species shows stronger reverse bias. We show how long range order is preserved in this case because of vacancies. Presence of vacancies also gives rise to a new kind of ordering where phase separation of one species of particles takes place on a plateau like structure of the landscape, \textcolor{black}{where the surface dynamics can be mapped to the partially asymmetric exclusion process with open boundary in the maximal current phase. Due to power law relaxation of the boundary layer, height of the plateau scales as $\sqrt{N}$, unlike macroscopic hills and valleys of the landscape seen earlier. This is the underlying mechanism responsible for the novel kind of phase ordering.} Our study highlights the important role played by vacancies: although the vacancies do not put any bias on the landscape, they can significantly influence the competition between aligned and reverse bias, and consequently ordering in the system.

The newly discovered VIPS phase shows strong finite size effects. As seen from Fig. \ref{fig:Hfrac} the phase separation is less and less complete for smaller $N$ values. Although the phase diagram shown in Fig. \ref{fig:phasev}(a) indicates that VIPS phase continues until both particle species show reverse bias (third quadrant of Fig. \ref{fig:phasev}(a)), this holds only for very large $N$. For moderate or small $N$, the long range order is lost even when one species shows aligned bias. We show this explicitly in Appendix \ref{sec:HFraction} by measuring subsystem density fluctuations of the phase separated species at different points in VIPS phase. By looking at how these density fluctuations scale with the subsystem size \cite{das2025persistent, dey2012spatial, kuroda2023anomalous, narayan2007long}, one can identify the boundary between the disordered and VIPS phases. Indeed we find for smaller $N$ the boundary is shifted from what is shown in Fig. \ref{fig:phasev}(a). To explain the origin of this rather strong finite size effect, we note that the height of the plateau is significantly smaller than macroscopic hills seen in other ordered phases. Therefore, the current fluctuations present in the flat segment of the landscape for small $N$ may destabilize the plateau and disperse the clusters. It will be useful to calculate a bound for $N$ below which this happens. The bound is expected to be a function of $b$ and $b'$. More research is needed in this direction.

Finally, a more generalized version of LHV model can be defined by introducing a relative time-scale between the surface movement and particle movement. This time-scale determines how fast (or slow) the surface moves compared to particles. A similar idea was implemented for LH model in \cite{khamrai2024effect}. In particular, by tuning this time-scale it is possible to ensure product measure in the disordered segment of FPPS or VIPS phase \cite{khamrai2024effect}. This will make it possible to quantitatively match our analytical predictions from mean field theory with the simulation results. The mismatch seen in Fig. \ref{fig:Hfrac} for instance, goes away in that case.  However, for the sake of simplicity, we have considered here equal time-scale for surface and particle movements. It is a simple exercise to generalize our results for the case when the two time-scales are different.

\section{Acknowledgements}
CK acknowledges research fellowship (Grant No. 09/0575(12571)/2021-EMR-I) from the Council of Scientific and Industrial Research (CSIR), India. SC acknowledges support from Anusandhan National Research Foundation (ANRF), India (Grant No: CRG/2023/000159).

\appendix

\section{Additional Data for FPPS Phase} \label{sec:FPPSPhase}

\begin{figure}[H]
\centering
\includegraphics[scale=1]{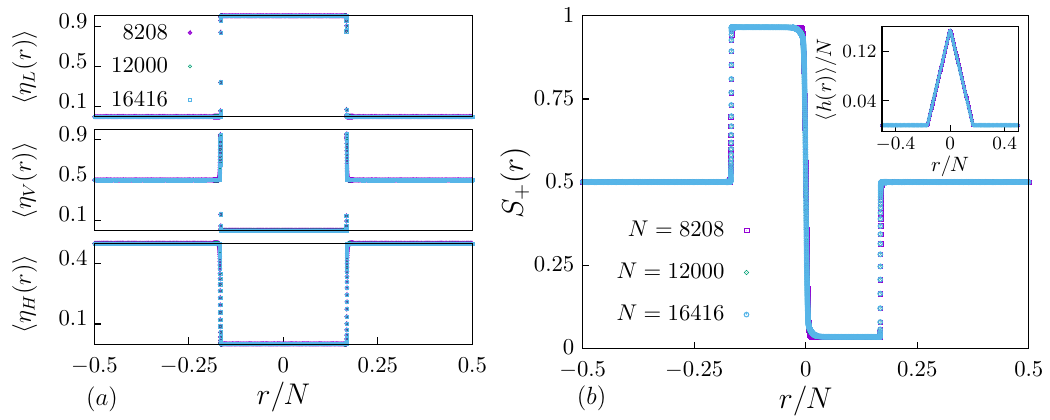}
\caption{(a) Particle occupancies as a function of scaled distance $r/N$. (b) Probability to find an upslope bond at a scaled distance $r/N$. The probability remains distinctly less than unity and above zero, which proves the existence of finite fraction of tilt impurity in the hill. Inset: Average height profile shows a macroscopic hill. Here $a = 0.4$, $b' = 0.4$ and $b = -0.1$. The densities of the particle species are $\rho_{H} = \rho_{L} = 1/3$.} \label{fig:FPPSOcc}
\end{figure}

In Fig. \ref{fig:FPPSOcc}(a) we plot the steady state density profile for $L$ particles, $H$ particles and vacancies. We measure all distances from the center of mass for $L$ particles. $\langle \eta_L(r) \rangle $ denotes the density of $L$ particles at a distance $r$ and it takes the value $1$ for $|r| < N_L/2$ and then sharply drops to zero for larger distance. At the boundary of the $L$ cluster, near the foot of the macroscopic hill we find a small patch where $\langle \eta_V(r) \rangle $ has a large value, {\sl i.e.} vacancies are present here with a large probability. For $r$ slightly larger than this, a homogeneous mixture of vacancies and $H$ particles is found.

In Fig. \ref{fig:FPPSOcc}(b) we plot the probability $S_+(r)$ of finding an upslope bond at a distance $r$ from the $L$-cluster center of mass. Consistent with the structure of a macroscopic hill, $S_+(r)$ shows a sharp transition from a high value close to $1$ to a low value close to $0$ across $r=0$. In the inset we show the average height profile of the landscape which shows a macroscopic hill. 

\section{Finite Size Effects in VIPS-disordered Phase Boundary}\label{sec:HFraction}

\begin{figure}[H]
\centering
\includegraphics[scale=1]{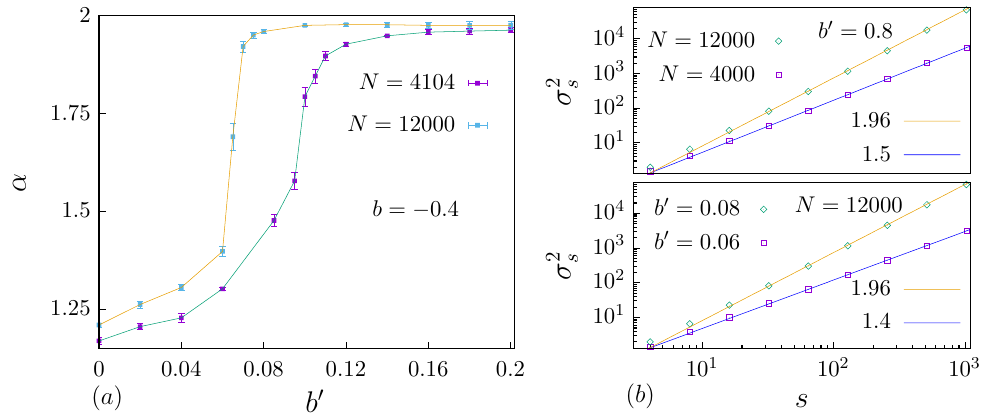}
\caption{(a) Variation of the exponent $\alpha$ with $b'$ for different $N$ and fixed $b=-0.4$. (b) Top panel shows subsystem density fluctuations $\sigma_s^2$ against subsystem size $s$ for fixed $b' = 0.8$ and different $N$. Bottom panel shows the same plot for fixed $N = 12000$ and different $b'$ values. The densities of the particle species are $\rho_{H} = \rho_{L} = 1/3$.} \label{fig:CoarCon}
\end{figure}

In VIPS phase one particle species phase separates but does not form a pure cluster. A small but finite fraction of the other species is present in it. This fraction becomes larger for smaller $N$ (see Fig. \ref{fig:Hfrac}). To check whether this trend affects the phase boundary between VIPS and disordered phase, we measure subsystem density fluctuations, defined as $\sigma_s ^2 = \langle [N_{L}(s) - \rho_{L}s]^2 \rangle$, where $N_{L}(s)$ denotes the number of $L$ particles in a subsystem of size $s$, which is placed at any arbitrary location on the landscape. We consider subsystem size $s$ in the range $1 \ll s \ll N$. If the system is in a disordered state, the $L$ particles are distributed homogeneously. Then from central limit theorem it follows that the variance $\sigma_s ^2 \sim s^\alpha$ with $\alpha =1$. On the other hand, if $L$ particles are in a phase separated state then macroscopic clusters are present and $N_L(s) =s$ within such clusters and $N_L(s) =0$ elsewhere. Other values of $N_L(s)$ occur with much smaller probability for  $1 \ll s \ll N$. For such states,  $\sigma_s ^2 \sim s^2$, {\sl i.e.} $\alpha=2$ \cite{das2025persistent, dey2012spatial, kuroda2023anomalous, narayan2007long}. In Fig. \ref{fig:CoarCon} we plot $\alpha$ as a function of $b'$ for a fixed $b < 0$. As expected, $\alpha$ approaches $2$ for large $b'$ and sharply decays to a value close to $1$ as $b'$ decreases. For smaller $N$ our data in Fig. \ref{fig:CoarCon} shows that the transition takes place at a higher $b'$ value, {\sl i.e.} the effect of the disordered phase is felt early on. For larger $N$, however, the transition point is shifted to lower $b'$. This trend is consistent with the phase diagram of LHV model shown in Fig. \ref{fig:phasev}(a) where transition takes place at $b'=0$. Finite size effects disappear rather slowly in this case and computationally it is not feasible for us to go to much larger $N$.

\end{document}